\documentclass[%
 reprint,
 amsmath,amssymb,
 aps,
]{revtex4-2}

\usepackage{graphicx}
\usepackage{dcolumn}
\usepackage{bm}
\usepackage[caption=false]{subfig}

\begin{document}

\preprint{APS/123-QED}

\title{Two-parameter landscape of transport efficiency in mesoscopic networks: transitions from the Braess to normal regimes without a congestion relaxation}

\author{A.D. Lobanov}
\email{lobanov.ad@phystech.edu}
\affiliation{ Moscow  Institute  of  Physics  and  Technology, 9  Institutskiy  per.,  Dolgoprudny,  Moscow  Region
141701, Russia.}
\author{A.D. Lobanova}%
\email{evstafeva.ad@phystech.edu}
\affiliation{ Moscow  Institute  of  Physics  and  Technology, 9  Institutskiy  per.,  Dolgoprudny,  Moscow  Region
141701, Russia.}
 \author{A.M. Pupasov-Maksimov}
 \email{tretiykon@yandex.ru}
\affiliation{ Universidade Federal de Juiz de Fora
 Department of Mathematics }%


%
%


\date{\today}

\begin{abstract}
This paper deals with the Braess paradox in quantum transport. The scattering matrix formalism is used to consider a two-parameter family of mesoscopic conductors with the topology of the classical Braess transport network. The study investigates the mutual influence of the congestion and smoothness of the system on the Braess behavior. Both the Braess paradox and normal transport regimes coexist within the two-parametric landscape under the same congestion.  
\end{abstract}

\maketitle


\section{\label{sec:level1}Introduction} 

The Braess paradox is an unusual behavior of a network in response to the addition of  "extra" paths. Instead of improvements in transport efficiency, "extra" paths impair network traffic. This phenomenon has been found in almost all fields of physics. Its presence is recorded in road transport networks, in electric circuits, hydraulic gas networks and even mechanical networks \cite{nash1950equilibrium,braess1968paradoxon,cohen1991paradoxical}. An analogue of this process in nuclear physics could be the nuclear Overhauser effect \cite{neuhaus2007nuclear}. Recently, such behaviour was predicted and observed experimentally in a mesoscopic conducting network \cite{pala2012new}. The effect implies a weakening of the current in an extended network contrary to an expected increase. 

This study considers theoretical results \cite{pala2012new,pala2012transport} within the framework of the scattering matrix approach, using its equivalence with the non-equilibrium Green functions method \cite{kloss2021tkwant}. Further, we extend the model \cite{pala2012new} to take into account other possible parameters of a mesoscopic network. It was shown \cite{pala2012new} that releasing congestion considerably relaxes the paradoxical behavior of the network. It happens because of allowing a larger number of conducting channels to propagate in the region inside the structure. Here we demonstrate that Braess behaviour can also be eliminated by smoothing the network whose branches still maintain a congested transport regime. Smooth connections correspond to adiabatic changes in structure width along the electron wave function propagation. We construct a landscape of 2-3 branched transport system properties according to two competing parameters -  adiabaticity of the system (smoothness) and congestion of additional branches. This construction could lead to a better understanding of quantum transport. 

Numerical calculations of the corresponding tight-binding model were done with the Kwant package of Python 3 \cite{groth2014kwant}. The paper is organized as follows. It starts with the theoretical implementation of the model and the reproduction of the main results of \cite{pala2012new,pala2012transport}. Then a smoothed version of branched network is introduced to provide calculations of the two-parameter landscape of transport properties. The main results are discussed in conclusion.

\begin{figure*}
\includegraphics[scale=0.4]{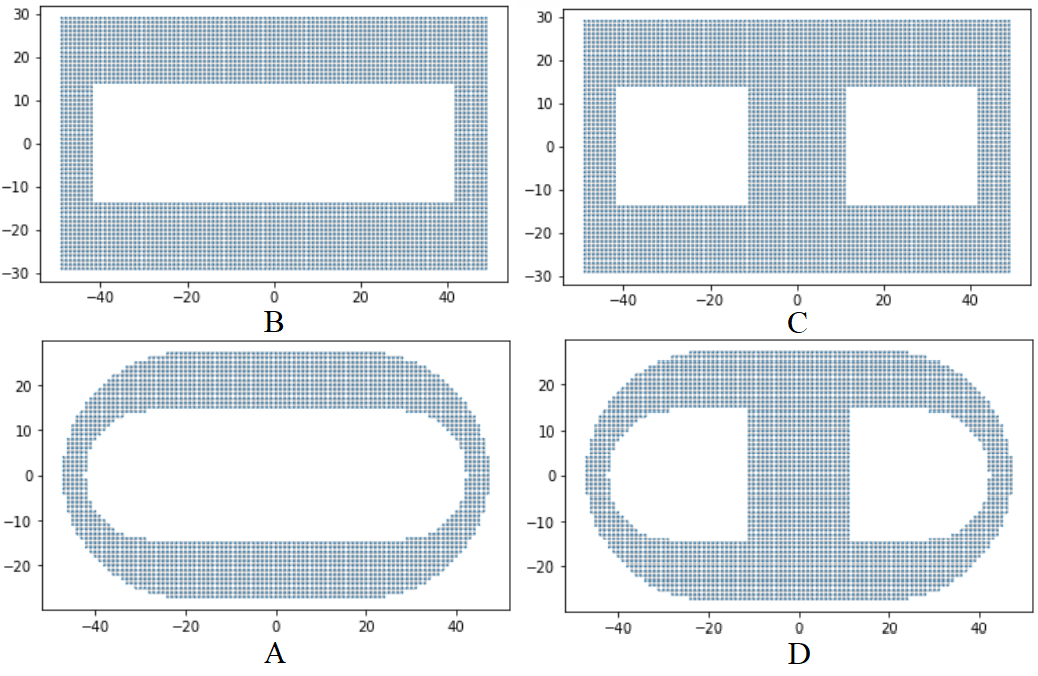}
\caption{We study 2-parametric family of mesoscopic systems. The first parameter is the width $W_3$ of the third path. The second parameter is responsible for continuous deformation from rectangular to ellipsoidal contour. The rectangular shape is equivalent to the system from \cite{pala2012new} in terms of geometry.}\label{fig:profiles}
\end{figure*}

\begin{figure*}
\includegraphics[scale=0.85]{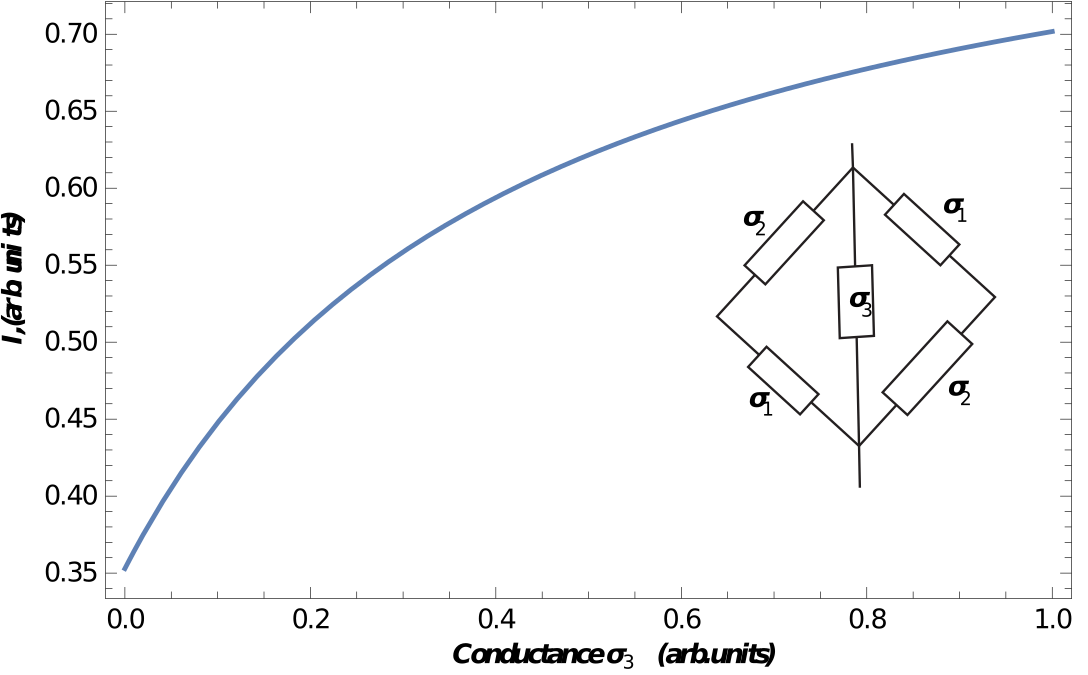}
\caption{Absence of any Braess behavior in the classical circuit without active elements or additional voltage drop. Increase of conductance $\sigma_3$ leads to the corresponding increase of current when the bridge is unbalanced ($\sigma_1\neq\sigma_2$). In the case of $\sigma_1=\sigma_2$ the current does not depend on $\sigma_3$.}\label{fig:classical-circuit}
\end{figure*}

\section{Theoretical implementation of the model}
Following \cite{pala2012new}, we compare transport properties of two mesoscopic samples with 2 and 3 ring structures (see Fig.~\ref{fig:profiles} B, C). In the case of an analogous classical electric circuit, Braess behavior is absent (see Fig.~\ref{fig:classical-circuit}) since no  active elements (Zener diodes) or additional voltage drops are added to simulate transportation costs analogous to the original transport network \cite{cohen1991paradoxical,nagurney2016observation}.  It is obvious that the difference between the quantum problem and the classical one is that the connection of two subsystems should be considered as a perturbation, and this leads to their mutual influence and a change in physical properties (spectrum, transmission coefficient, etc.). Note that  in classical networks (electrical, hydraulic), connections between elements can also perturb the properties of the system; for example, the non-zero hydraulic resistance of the tee or the bending of the pipe. In \cite{case2019braess} such non-linearity was used to cause the Braess paradox in microfluidic networks. 

For computational purposes we downscale the linear size of the sample from \cite{pala2012new} by factor 10. The basis network takes a rectangular form with approximately 90 sites in horizontal direction and 60 sites in vertical direction. It is connected to bottom (source) and top (drain) ohmic leads via two planar wires (openings) of width $W_{0} = 20$ sites. The left and right lateral wires of width $W_{1} = W_{2} = 10$ sites are narrower than the top and bottom horizontal paths. Narrow lateral paths ensure congested transport mode of the electrons propagation. The central wire directly connects the source to the drain and creates a third alternative path of propagation (Fig.~\ref{fig:profiles}).

Using an analogy with hydrodynamics we also considered an analogue of smooth bends in hydraulic networks to test its influence on the propagation of transverse modes along the conductor. Notably, the width of the lateral wires stays unchanged to maintain the congested regime which is necessary for the Braess paradox. 

The geometry of our network is defined 
by the following system of inequalities (see also code \cite{code})
\begin{eqnarray}
1.2 x_l^n + 1.2 y_l^n < 30^n\quad  \text{and}\quad  x_s^n + y_s^n > 14^n \cr \text{or}\quad |x|<W_3 \quad \text{and} \quad |y|<25.5\,,
\end{eqnarray}
where 
$$ x_l = \max(|x| - 20, 0),\qquad x_s= \max(|x| - 28, 0).$$
Here, to modify the geometry of our network from Fig.~\ref{fig:profiles}, B, C to Fig.~\ref{fig:profiles}, A, D, we introduce parameter $n$  which defines the geometry of external and internal boundaries of lateral wires. Thus, the width of the central path $W_3$ and the roundness parameter $n$ are used to define two-parametric family of mesoscopic networks.

\begin{figure*}
\includegraphics[scale=0.5]{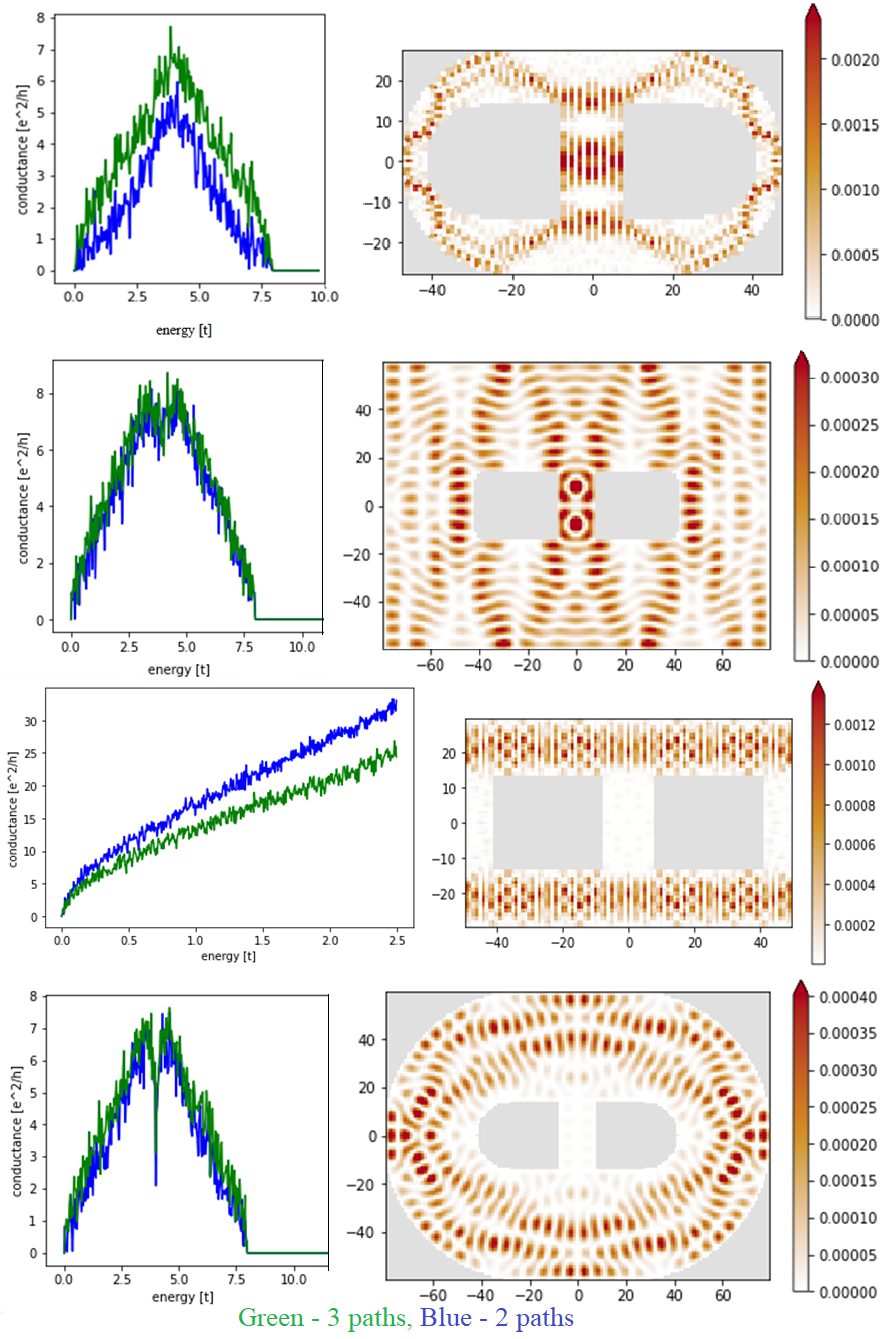}
\caption{The dependence of conductivity on energy for the different profiles. Here it is clearly seen that the rounding removes any hint of the presence of Braess paradoxical behavior.}\label{fig:everyprofileconductance}
\end{figure*}

\begin{figure*}
\includegraphics[scale=0.5]{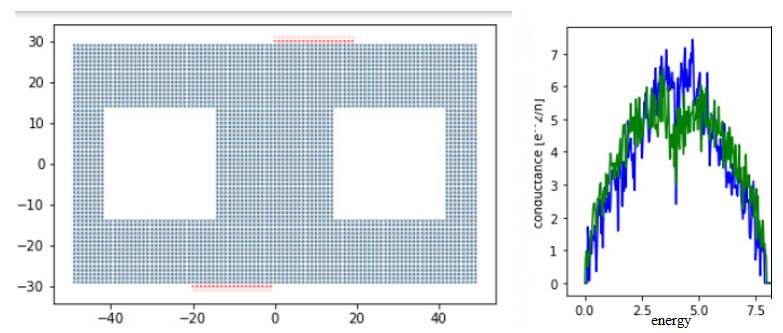}
\caption{The profile with no "congestion" state.}\label{fig:no-congest}
\end{figure*}

To detect the Braess-like behavior, it is sufficient to compare the total current through the systems, varying the width $W_3$ of the third path.
Therefore, we can use Landauer-Büttiker Formalism based on the scattering matrix method \cite{lesovik2011,moskalets2011scattering}.
There are only two leads, and the current through the lead "a" reads
\begin{equation}
I_a=\frac{e}{h}\int dE |S_{ab}(E)|^2\left(f_b(E)-f_a(E)\right)\,,    
\end{equation}
where $S_{ab}(E)$ is the S-matrix,
and   
\begin{equation}
f_{a,b}(E)=
\frac{1}{1+\exp\left(\frac{E-\mu_{a,b}}{k_BT_{a,b}}\right)}\,, 
\end{equation}
are Fermi distribution functions in the reservoirs "a, b". 
Under the DC bias, $\Delta U_{ab}=U_a-U_b$, reservoirs "a" and "b" have different electrochemical potentials,
\begin{equation}
\mu_{a,b}=\mu_0+eU_{a,b}
\end{equation}
This approach is equivalent to the method based on non-equilibrium Green's functions  \cite{moskalets2011scattering,gaury2014numerical,kloss2021tkwant} 
\begin{equation}
G_{i,j}^R(t,t')=-i\Theta(t-t')\sum_\alpha\int \frac {dE} {2\pi} \psi_{\alpha E}(t,i)\psi_{\alpha E}^*(t',j)
\end{equation}
\begin{equation}
G_{i,j}^<(t,t')=i\sum_\alpha\int \frac {dE} {2\pi} f_{\alpha}(E)\psi_{\alpha E}(t,i)\psi_{\alpha E}^*(t',j)
\end{equation}
which were used in \cite{pala2012new}. Roughly speaking, one can solve the Dyson equation for the wave function, instead of the Green function. From the wave function we can extract scattering amplitudes and define all transport properties of the system. Moreover, numerical calculations of wave functions require less memory and operations. 

The graph of the dependence of conductivity on energy (Fig.~\ref{fig:everyprofileconductance}) clearly shows that the conductivity of the two paths is higher; moreover, the difference between them increases depending on the energy. It is also worth adding that if you continue the graph to large values, you can see that the difference in conductivities increases to a certain peak. Then, at this peak there is a sharp drop in conductivity for the two paths, after which the results are mirrored. This can be clearly seen in the models below. The results of the probability density distribution for a given mode are shown next to it (Fig.~\ref{fig:everyprofileconductance}).
In fact, the last result (Fig.~\ref{fig:everyprofileconductance}) varies greatly at different energies. It gives us grounds to assume that a sharp increase in conductivity for three paths simultaneously with a sharp drop for two is due to the fact that the electrons "go better" in the model with three paths for a given energy, and "worse" in the two-path model.

\begin{figure*}
\includegraphics[scale=0.45]{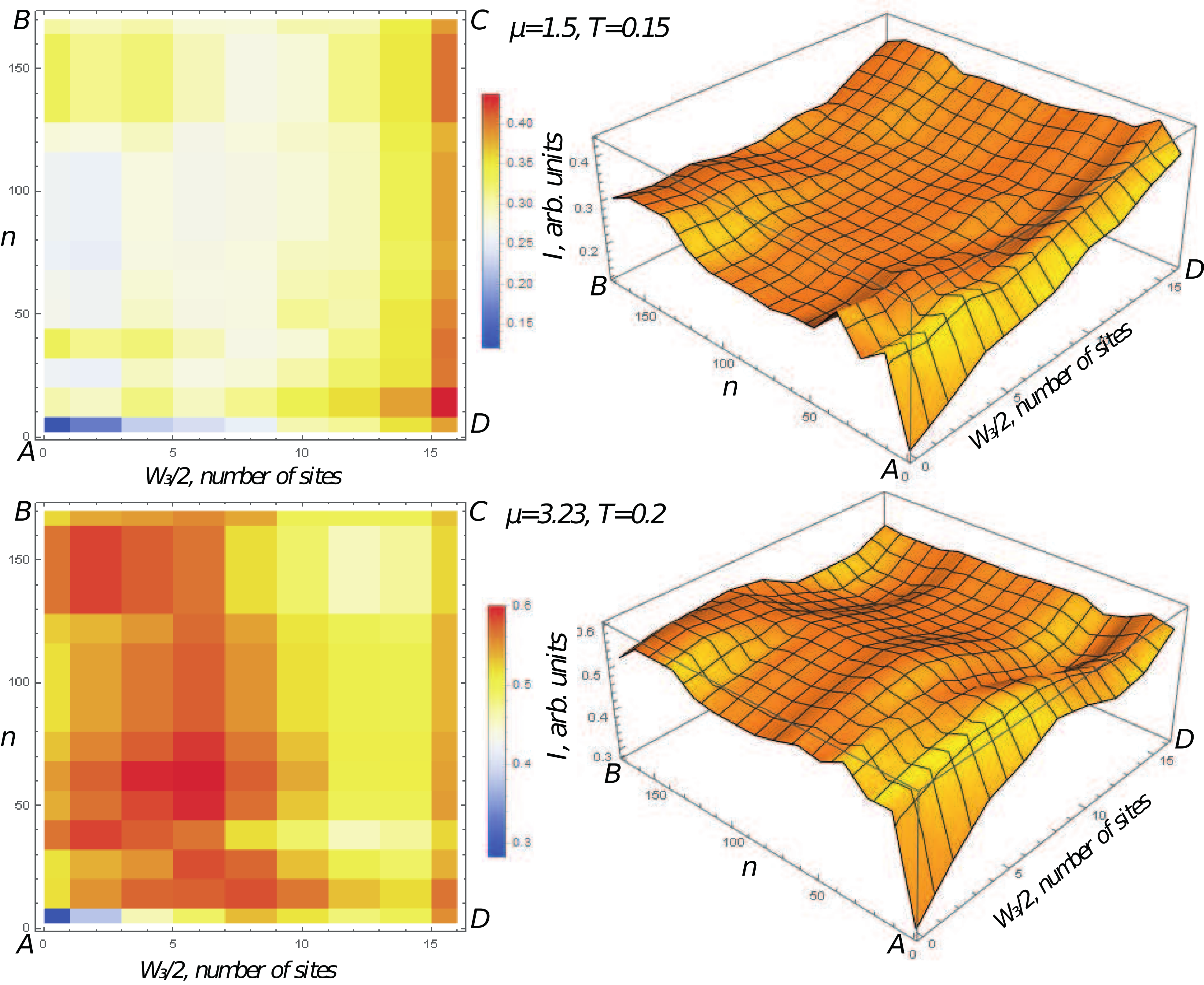} 
\caption{3D dependence of the total current on the width of the third channel and adiabaticity.}\label{fig:epsart}
\end{figure*}

\section{Two-parametric landscape of transport efficiency}
In this section we investigate the interplay between parameters $W_3$ and $n$ which results in a transition between the Braess paradox and normal transport regime without relaxing the congestion of the network.  

Below we first present some particular cases and plot the conductivity, as well as the probability density distribution for a given mode (Fig.~\ref{fig:everyprofileconductance}).

The results are dramatically different. The Braess behavior of the system has disappeared. Also at all energies it can be seen that, as intuitively assumed, a system with three paths wins in conductivity, relative to another system with two paths. The picture of the probability density distribution shows that the middle path is actively "used" by electrons. However, the total current here decreases three fold. There is no Braess behavior, and the new path improves conductivity, but the value of the total current drops. The same effect will be observed below, with other "rounded" geometries.

Thus, it can be said that adiabaticity affects the conductivity of the system. However, it is worth checking whether there is a place for the concept of "congestion" in such a system, or whether such an effect is associated only with adiabaticity.

To this end, we reproduced another result of  \cite{pala2012new}, where all paths widen to a state where there is no "congestion" (Fig.~\ref{fig:no-congest}), (Fig.~\ref{fig:epsart}).

As it was shown in \cite{pala2012new}, the "broadening" of all system parameters, except for the middle path, leads to the conductivity of three paths being almost always higher than the conductivity of two. There are only several random peaks, where the conductivity for the two paths is higher. These peaks could be explained by fluctuations.

The resulting graph of the dependence of conductivity on energy is almost identical, with the only difference being that the "rounded" version has a stronger negative peak, as here (Fig.~\ref{fig:epsart}). This means that adiabaticity affects the results, but it is not the only reason for such Braess-like behavior. The probability density distribution shows that the middle path is almost not "congested". These results lead to the idea that the concept of "congestion" is not unambiguous.

To show that it is not only the demand that affects here, we can give an example of a situation where the source and drain are smaller than the central wire (Fig.~\ref{fig:epsart}) Here, the demand will be minimal, however, according to the energy conduction graph, there are still energies with Braess behavior.

Finally, we construct a landscape of transport efficiency of the considered network. We plot a 3-D graph of the total current as a function of the width of the central wire $W_3$ and the roundness parameter $n$ on Fig.~\ref{fig:epsart} . To test stability of the results we repeat calculations at different temperatures and electro-chemical potentials.

It can be seen that in the case of a smooth (rounded) system, the total current is increased as a function of $W_3$ thus avoiding the Braess paradox. Moreover, increasing $n$ we obtain regions with Braess and normal behaviour as a trivial consequence of a non-flat landscape. 

\section{Conclusion}
To detect Braess-like behavior it is sufficient to calculate the total current through the system, therefore the scattering matrix approach can be efficient. Since the solution of  Dyson equation for the non-equilibrium Green function requires large computational resources, it is difficult to study transport properties of a system under variation of several parameters. In our work we demonstrated how such a study can be done with an alternative approach. 
First of all, our two-parametric family  of mesoscopic systems contains a region already studied in \cite{pala2012new,pala2012transport}. It is a region close to the BC segment of our parameter space (see Fig.~\ref{fig:epsart}). The behaviour of our system in this region is in accordance with the results of \cite{pala2012new}. In \cite{pala2012new} robustness of Braess-like behaviour was proved by considering a large sample with many conducting channels contributing to the total current and by varying the Fermi wavelength.  Our calculations also confirm that this Braess-like behaviour is not related to phase-coherent or size effects. However, we also found regions in the parameter space, where a mesoscopic sample with the same geometrical congestion of lateral branches does not demonstrate Braess-like behavior. Indeed, the total current increases along the AD segment in the upper part of Fig.~\ref{fig:epsart}. Therefore we can conclude, that the Braess-like behavior observed in \cite{pala2012new,pala2012transport} is not related to geometrical congestion of the network. 

Actually, it is a consequence of a nonlinear landscape of the total current as a function of parameters of the model. Let us fix a point of the parameter space and consider the corresponding Hamiltonian and wave functions unperturbed. Small changes of parameters (width $W_3$ or smoothness of boundaries) can be considered as a perturbation. Total current is the expectation value of the current operator, therefore its derivative with respect to the perturbation parameter can be calculated by the finite temperature version of Hellmann–Feynman theorem \cite{pons2020hellmann}. In general, the sign of this derivative can be arbitrary depending on the model. Only in the case of an electrical current passing through a perfectly smooth narrow constriction will such a derivative be non-positive. This is because the current is carried by a finite number of quantized modes (analogous to those in a waveguide), each of which contributes $2e^2/ h$ to the conductance \cite{krans1995signature} the maximum current is achieved. As a result, any perturbation will result in the decrease of the total current. This decrease may correspond to Braess-like behavior when one increases the width of conductor thus adding conducting channels, while the source and drain leads stay unchanged.     

In a more complicated model with 2-ring geometry we have shown that one can relax Braess behavior without relaxing geometrical congestion of the network. Finally, we can speculate that observing a 3-parameter family of mesoscopic systems, one may find typical singular points in the level sets of the total current landscape which lead to bifurcations in system behaviour and jumps in its characteristics. These findings have the potential to advance the development of built-in control mechanisms in mesoscopic networks.    

\nocite{*}

\bibliography{libr}

\end{document}